\documentclass[fleqn,10pt]{wlscirep}

\usepackage{multirow}

\title{Micro Fourier Transform Profilometry ($\mu$FTP): 3D shape measurement at 10,000 frames per second}

\author[1,2,*]{Chao Zuo}
\author[1,2]{Tianyang Tao}
\author[1,2]{Shijie Feng}
\author[3]{Lei Huang}
\author[4]{Anand Asundi}
\author[2,*]{Qian Chen}
\affil[1]{Smart Computational Imaging (SCI) Laboratory, Nanjing University of Science and Technology, Nanjing, Jiangsu Province 210094, China}
\affil[2]{Jiangsu Key Laboratory of Spectral Imaging $\&$ Intelligent Sense, Nanjing University of Science and Technology, Nanjing, Jiangsu Province 210094, China}
\affil[3]{Brookhaven National Laboratory, NSLS II 50 Rutherford Drive, Upton, NY 11973-5000, United States}
\affil[4]{Centre for Optical and Laser Engineering (COLE), School of Mechanical and Aerospace Engineering, Nanyang Technological University, Singapore 639798, Singapore}

\affil[*]{Correspondence and requests for materials should be addressed to C.Z. (email: zuochao@njust.edu.cn) or Q.C. (email: chenqian@njust.edu.cn)}

\begin{abstract}
Recent advances in imaging sensors and digital light projection technology have facilitated a rapid progress in 3D optical sensing, enabling 3D surfaces of complex-shaped objects to be captured with improved resolution and accuracy. However, due to the large number of projection patterns required for phase recovery and disambiguation, the maximum fame rates of current 3D shape measurement techniques are still limited to the range of hundreds of frames per second (fps). Here, we demonstrate a new 3D dynamic imaging technique, Micro Fourier Transform Profilometry ($\mu$FTP), which can capture 3D surfaces of transient events at up to 10,000 fps based on our newly developed high-speed fringe projection system. Compared with existing techniques, $\mu$FTP has the prominent advantage of recovering an accurate, unambiguous, and dense 3D point cloud with only two projected patterns. Furthermore, the phase information is encoded within a single high-frequency fringe image, thereby allowing motion-artifact-free reconstruction of transient events with temporal resolution of 50 microseconds. To show $\mu$FTP’s broad utility, we use it to reconstruct 3D videos of 4 transient scenes: vibrating cantilevers, rotating fan blades, bullet fired from a toy gun, and balloon’s explosion triggered by a flying dart, which were previously difficult or even unable to be captured with conventional approaches.
\end{abstract}
\begin{document}

\flushbottom
\maketitle
%
%
\thispagestyle{empty}

\section*{Introduction}

\noindent
The desire to capture and record fast phenomena can be traced back to invention of film photography in the nineteenth century, which used controllable mechanical light shutters to limit the film exposure time
\cite{1,2,3}. However, it is only with the advances made during the last few decades in the field of solid-state electronic imaging sensors based on the charge-coupled device (CCD) or complementary metal-oxide-semiconductor (CMOS) technology that high-speed imaging has gained substantial interests and applications \cite{4,5}. Investigation into the scientific origins of fast phenomena has benefited enormously from the development of such high-speed cameras \cite{6,7}, and other applications exist in almost all areas of industry, television, defense, sports, and health care. More recently, ultra-high-speed imaging at picosecond (10$^{-12}$s) timescales has been demonstrated by either using ultra-short pulse illumination to provide temporal resolution \cite{8,9}, or combining streak imaging technology with scanning \cite{10} or temporal pixel coding strategy \cite{11} to achieve two-dimensional (2D) information, enabling a frame rate fast enough to visualize photons in motion.

Despite these tremendous advances, most of the current high-speed imaging sensors can only record instantaneous phenomena as 2D image sequences that lack the depth information. Nevertheless, high-speed three-dimensional (3D) reconstruction of transient scenes has been long sought in a host of fields that include biomechanics \cite{12}, industrial inspection \cite{13}, solid mechanics \cite{14}, and vehicle impact testing \cite{15}. In these areas, it is always desirable that the 3D information can be acquired at a high frame rate so that transient geometric changes of an object or an environment can be captured. These can then be reviewed in slow motion to provide in-depth insights into fast changing events in a broad range of application scenarios.

Over the past decades, 3D image acquisition technologies have also rapidly evolved, benefiting from the advances in electronic imaging sensors, optical engineering, and computer vision. For a small scale depth or shape, micrometer or even nanometer measurements can be reached by using interferometry \cite{16}, confocal \cite{17} or other depth-resolved 3D microscopic techniques \cite{18,19}. Time-of-flight methods \cite{20,21}, in which the distance is resolved by measuring the time of flight of a light pulse with the known speed of light, are well suited for measuring large-scale scenes or depth ($>$ 0.5 m). For the 3D measurement of medium-size objects, full field triangulation based active structured light (SL), has proven to be one of the most promising techniques \cite{22,23,24}. Just as human vision system, the basic principle of SL is optical triangulation, in which correspondences between a projector and a camera are established with some coded patterns projected onto the scene. Recent years, numerous SL approaches have been proposed and investigated, and there is a clear trend towards improving the measurement speed to video rates ($\sim$ 25 Hz) and far beyond \cite{25,26,27}. This trend is being driven by the increasing demand for high-speed depth data coupled with rapid advances in high-frame-rate image sensors and digital projection technology. Today’s high-speed cameras can capture video at speeds up to tens of thousands full frames per second (fps) and even faster at a reduced resolution. On the other hand, the digital micro-mirror device (DMD), as a key component of digital light processing (DLP) projection system, is able to create, store, and display high-speed binary pattern sequences through optical switching ( ``on" and ``off" ) at rates in excess of 10,000 Hz as well. By operating the DMD in binary (1-bit) mode, quasi-sinusoidal fringe patterns can be created at the maximum frame rates with lens defocusing and binary dithering techniques \cite{28,29,30}.

Once the speed of hardware is no longer a limiting factor, the main hurdle to overcome in high-speed 3D sensing is to reduce the number of patterns required for reliable 3D reconstruction. Single-shot approaches \cite{31,32}, e.g., de Bruijin sequences \cite{32,33}, M-arrays \cite{34}, and symbol coded patterns \cite{35}, are well-suited for dynamic 3D sensing. Generally, the pattern needs to be wisely designed so that each point can be uniquely identified from its neighboring pixels. Nevertheless, the spatial resolution and depth accuracy of single-shot approaches are limited due to the local depth smoothness assumption, which does not hold around the regions with depth discontinuities and fine details. Fourier transform profilometry (FTP) \cite{25,36,37,38} is another representative single-shot approach in which the phase is extracted from a single high-frequency fringe image by applying a properly designed band-pass filter in the frequency domain. Compared with other single-shot methods, FTP has the advantages of pixel-wise measurement and effective noise removal, yet with the precondition that the fundamental frequency component, which carries the phase information of the object, is well separated with zero-order background. In practice, this precondition can be easily violated when the measured surface contains sharp edges, discontinuities, and/or large surface reflectivity variations \cite{37,39,40,41,42}. Several methods have been proposed to alleviate this problem by carefully designing the band-pass filter \cite{40,43}, or capturing an additional flat \cite{41} or  $\pi$-phase shifted fringe pattern \cite{39}. However, for objects with complex surface properties, the measurement accuracy of these FTP approaches is still quite limited due to the inherent spectrum leakage.

To achieve high-quality dense 3D reconstructions, the multi-frame methods \cite{44,45,46} are usually preferred for their advantages of high accuracy and low complexity. In general, multi-frame methods require several predetermined patterns to be projected onto the measured object. Hence, their measurement speed is limited by the number of patterns per sequence and both camera and projector frame rate. The most widely used multi-frame approach is phase shifting profilometry (PSP) \cite{47}, which requires a minimum of three fringe images to provide high-accuracy pixel-wise phase measurement. Furthermore, the measurement is quite robust to ambient illumination and varying surface properties. However, when measuring dynamic scenes, motion will lead to phase distortion artifacts, especially when the object motion during the inter-frame time gap is non-negligible \cite{42,46,48,49,50}. This is an intrinsic problem of PSP, since the phase information is spread over multiple fringe images. Another challenging problem in both PSP and FTP is the phase ambiguity resulting from the periodical nature of the sinusoidal signal. Though high-frequency patterns with dense fringes are usually preferred for high-accuracy 3D reconstruction (especially for FTP), they also introduce severe ambiguities. To recover the absolute phase, a common practice is to use temporal phase unwrapping (TPU) algorithms with the help of Gray-code patterns \cite{51} or multi-wavelength fringes \cite{52}. However, this requires a large number of additional images, which are only used for phase disambiguation purpose but not contribute to the final 3D results. The prolonged pattern sequence (e.g., a minimum of 9 patterns are required per 3D reconstruction for three-wavelength PSP \cite{53}) greatly limits the performance of PSP and FTP in high-speed, time-critical scenarios.

To address this limitation, several composite phase shifting schemes \cite{54,55,56,57,58,59} (e.g., dual-frequency PSP \cite{54}, bi-frequency PSP \cite{59}, 2+2 PSP \cite{55}, and period coded PSP \cite{56,57}) have been proposed. They can solve the phase ambiguity problem without significantly increasing the number of projected patterns. However, in order to guarantee a reasonable reliability of phase unwrapping, the fringe frequency often cannot be too high, which provides only a comparatively low accuracy \cite{60}. Alternatively, stereo unwrapping methods based on geometric constraint can be used to determine the fringe order without capturing additional images \cite{46,58,61,62,63}. But the measured objects must be within a restricted depth of the measurement volume. Moreover, some of these techniques require multiple ($\geq$2) high-speed cameras, which could considerably increase the overall cost of the system. On a different note, high-speed 3D sensing can also be realized via active stereo-photogrammetry without explicit evaluation of phase information \cite{64,65,66,67,68}. In such approaches, the projected patterns are merely used to establish precise camera-to-camera correspondences based on correlation algorithms, so various types of structured patterns (e.g., statistical speckle \cite{64} and aperiodic sinusoid \cite{69}) can be applied. This idea allows to develop alternative projection units without the need of a DMD, yielding simpler optical designs and/or much higher projection rates. For example, the LED-based array projection system and the GOBO projection system enable high-speed pattern switching with frame rates up to tens of kHz \cite{66,67}, which is much higher than the maximum speed of commercial DLP projectors (typically 180-360 Hz after disassembling the color wheel \cite{45,55,70}). However, these techniques still require at least two high-speed cameras. Moreover, as stated by Grosse \emph{et.al.} \cite{71}, more than 9 images are typically required to establish dense, accurate, and outlier-free correspondences. Thus, the final 3D frame rate achievable is still far lower (almost an order of magnitude) than the native frame rate of the camera and projector, which is typically in the range of only several hundreds Hz (e.g., 330 Hz at a projection frame rate of 3 kHz by using array projection \cite{66} and 1,333 Hz at a camera rate of 12,000 Hz by using GOBO projection \cite{67}).

To overcome above limitations, here we present \emph{Micro Fourier Transform Profilometry} ($\mu$FTP), which enables highly-accurate dense 3D shape reconstruction at 10,000 Hz, without posing restrictions on surface texture, scene complexity, and object motion. In contrast to conventional FTP which uses a single high-frequency sinusoidal pattern with a fixed spatial frequency, $\mu$FTP introduces very small temporal variations in the frequency of multiple spatial sinusoids to eliminate phase ambiguity. So in $\mu$FTP, the word \emph{Micro} just refers to the small values for both the temporal frequency variations and periods of fringe patterns, which is similar to the case of \emph{Micro} Phase Shifting \cite{72} but the latter uses band-limited high-frequency fringes to reduce the phase errors due to global illumination. The key idea of $\mu$FTP is to freeze the high-speed motion by encoding the phase information within a single high-frequency sinusoidal pattern. And the phase ambiguities are resolved spatio-temporally with the extra information from the small frequency variations along the temporal dimension. Besides, high-quality 2D texture can be acquired by inserting additional white patterns (all mirrors of the DMD are in the ``on" state) between each high-frequency sinusoidal patterns, which also remove spectrum overlapping and enable high-accuracy phase measurements even in the presence of large surface reflectivity variations. Unlike previous approaches in which the phase retrieval and disambiguation were separately addressed in a pixel-wise and time-dependent fashion, the $\mu$FTP extends phase unwrapping into the space-time domain. The main contributions of this paper are three-fold:

(1) We develop a high-frame-rate DLP fringe projection system that enables binary pattern switching and precisely synchronized image capture at a frame rate up to 20,000 Hz, which is faster than the previously reported setups. It is composed of a high-speed CMOS camera and a high-speed projection system based on a DLP development kit (Section \textbf{Experimental setup}). Technical details about the hardware setting and the optics module are discussed in Sections \textbf{Projection and capture synchronization} and \textbf{Projection optics}.

(2) We propose and analyze a complete computational framework (including phase recovery, phase unwrapping, error compensation, and system calibration) that allows to effectively recover an accurate, unambiguous, and distortion-free 3D point cloud with every two projected patterns (Section \textbf{Principle of $\mu$FTP}). Each projected pattern serves the dual purpose of phase disambiguation and 3D reconstruction, which allows to significantly reduce the number of projected patterns. The comprehensive theory, implementation, and demonstration of each algorithm involved in the $\mu$FTP framework are provided in details in \textbf{Supplementary Information}.

(3) By applying the $\mu$FTP framework to the high-frame-rate fringe projection hardware, we achieve an unprecedented 3D imaging speed up to 10,000 fps, with a depth accuracy better than 80 $\mu$m and a temporal uncertainties below 75 $\mu$m according to a large measurement volume of 400 mm $\times$ 275 mm $\times$ 400 mm (Section \textbf{Quantitative analysis of 3D reconstruction accuracy}). In Section \textbf{Results}, we demonstrate for the first time high-quality textured 3D imaging of vibrating cantilevers, rotating fan blades, flying bullet, and bursting balloon, which were previously difficult or even unable to be captured with conventional approaches.

\section*{Materials and methods}

\subsection*{Experimental setup}

\noindent
The high-frame-rate fringe projection system consists of a high-speed CMOS camera (Vision Research Phantom V611) and a high-speed projection system. The high-speed projection system includes a DLP development kit (Texas Instruments DLP Discovery 4100) with an XGA resolution (1024 $\times$ 768) DMD, and a custom-designed optics module. By omitting any grayscale capabilities, we drive the DMD at a refresh rate up to 20,000 binary fps. The light source is a green LED module with an output of 600 lumens. The high-speed camera used in this system has a frame rate of 6246 fps for maximum image resolution (1280 $\times$ 800). A 24 mm (focal length) lens (Nikon AF-S) with a variable aperture from f/3.5 to f/5.6 is attached to the camera. The camera lens aperture is fully opened to allow the maximum amount of light to enter. In this work, the camera is operated at a reduced image resolution (640 $\times$ 440) to match the frame rate of the DMD (20,000 fps) with an exposure time of 46 $\mu$s. It is also precisely synchronized with the projection system with the help of the DLP development hardware and a self-developed voltage translation circuit, which will be described in details in the next subsection.

\subsection*{Projection and capture synchronization}

\noindent
To realize high-frame-rate 3D shape reconstruction, the key technical issue regarding the hardware system is the precisely synchronized high-speed pattern projection and capture. Over the years, a number of fringe projection systems have been developed by re-engineering off-the-shelf DLP projectors for high-speed applications \cite{44,45,55,70}. The common idea is to remove the color wheel of the DLP projector to make it work in the monochrome mode so that the projection rate can be tripled theoretically. For conventional 3-step PSP, one can simply make the projector display one static color image with three phase-shifted patterns encoded in its RGB channels respectively \cite{44,45}. Once the projector receives the video signal, it will automatically decode the input color image and project the three phase-shifted patterns in each color channel sequentially. The situation becomes more complicated when the pattern sequence contains more than 3 images because the input color image needs to be changed sequentially and periodically at high speed. This can be accomplished by either introducing additional Field-Programmable Gate Array (FPGA) hardware for video signal generation \cite{55}, or directly manipulating the memory of the graphical device based on CUDA programming \cite{70}. Nevertheless, the maximum projection rate can be achieved is still around 180-360 Hz, which is ultimately limited by the projection mechanism of the DLP projector (the gray-scale pattern is generated with the binary pulse width modulation along the time-axis, and the intensity level is reproduced by integrating smaller binary intensities over time by an image sensor, like eye or camera) and the use of standard (typically 60 Hz-120 Hz) graphic adapters for DMD control. Such bottlenecks in speed block a regular consumer DLP projector from many important applications where very fast motion needs to be acquired.

\begin{figure}[ht]
\centering
\includegraphics[width=1.0\linewidth]{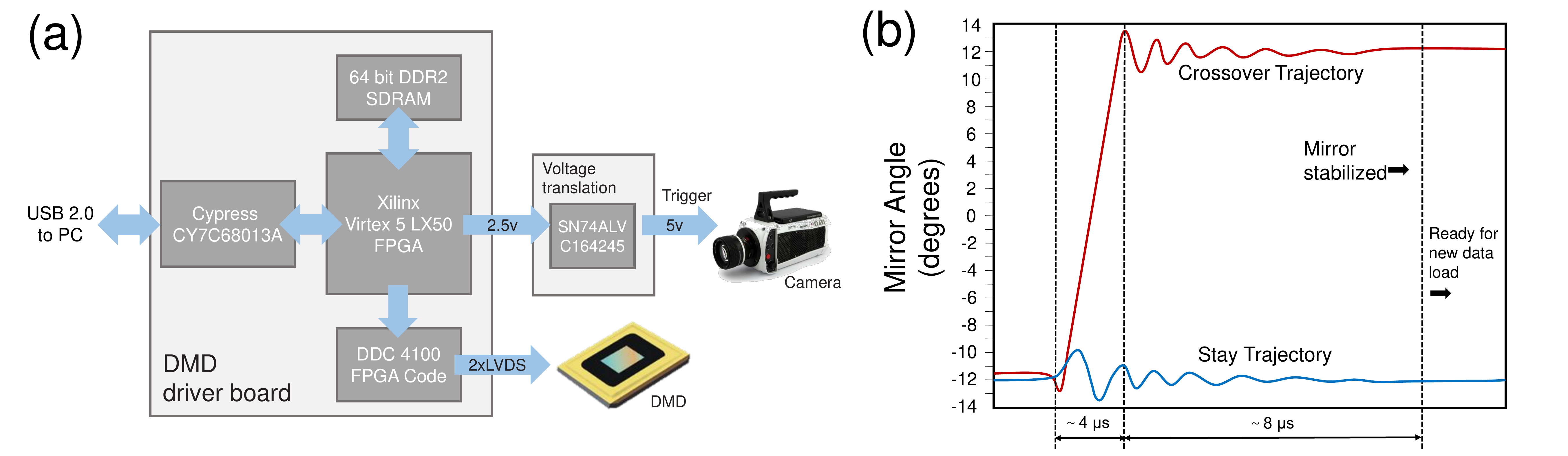}
\caption{(a) Schematic diagram of the high-frame-rate fringe projection system developed. (b) Mirror trajectory for the DMD after ``reset" operation.}
\label{fig1}
\end{figure}

To overcome these current limitations, in this work, we use a DLP Discovery 4100 development kit from Texas Instruments as the basis of our projection system. As shown in the hardware diagram [Fig. \ref{fig1}(a)], the DLP kit includes a DMD chip and a DMD driver board. The DMD is based on the 0.7” XGA (1024 $\times$ 768) chipset with a mirror pitch of 13.6 $\mu$m (DLP7000, Texas Instruments). The driver board has a programmable FPGA (Xilinx Virtex 5 LX50) and a USB 2.0 interface (Cypress CY7C68013A) for receiving input patterns from the computer. The on-board DDR2 SDRAM memory is used to store the pattern sequences that are pre-loaded for subsequent high-speed projection. The FPGA connects with the DMD controller chip (DDC4100, Texas Instruments) by parallel interface and transfers pattern data to it directly. At the same time, DDC4100 controls the mirrors of DMD to turn for generating measurement matrices according to the pattern data via a 2xLVDS interface. Based on these specific hardware, DLP Discovery 4100 offers advanced micro-mirror control as well as flexible formatting and sequencing light patterns. It enables the DMD to operate in binary mode without any temporal dithering, allowing binary light patterns to be projected with speed far surpassing that of a regular DLP projector. However, there remain a few constraints underlying the basic operation of a DMD.

(1) There hardware requires a certain amount of time to transition from one micro-mirror configuration to another - a limitation that is imposed due to data transmission to the DMD. The data transmission bus between FPGA and DMD operates at 64 bits and at 400 MHz. Hence, for our DMD with 1024 $\times$ 768 micro-mirrors, it takes at least 30.72 $\mu$s (${\tau _{LD}}$) to load a full frame binary image
\begin{equation}\label{1}
{\tau _{LD}}{\rm{ = }}\frac{{1024 \times 768}}{{64 \times 400 \times {{10}^6}}} = 30.72{\kern 1pt} {\kern 1pt} {\rm{\mu s}}
\end{equation}

(2) When the data is loaded in memory, ``reset" operation can be asserted, which tilts the mirrors into their new ``on" or ``off" states. It requires some time to physically tilt the mirrors and some time for mirrors to settle down. As illustrated in Fig. \ref{fig1}(b), there are about 4 $\mu$s transition time (${\tau _{TT}}$) and 8 $\mu$s settling time (${\tau _{ST}}$) during the mirror state conversion process. Besides, during the reset and mirror settling time, no data can be loaded in the DMD, thus slowing down the pattern output rate.

\begin{figure}[ht]
\centering
\includegraphics[width=1.0\linewidth]{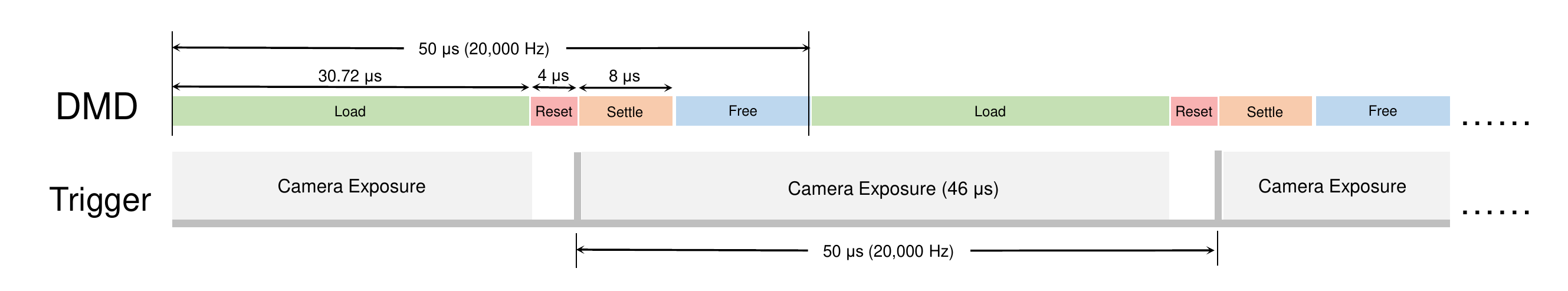}
\caption{Timing diagram of the designed hardware system. The exposure time of the camera is set at 46 $\mu$s to skip the mirror transition stage.}
\label{fig2}
\end{figure}

\begin{figure}[htb!]
\centering
\includegraphics[width=0.8\linewidth]{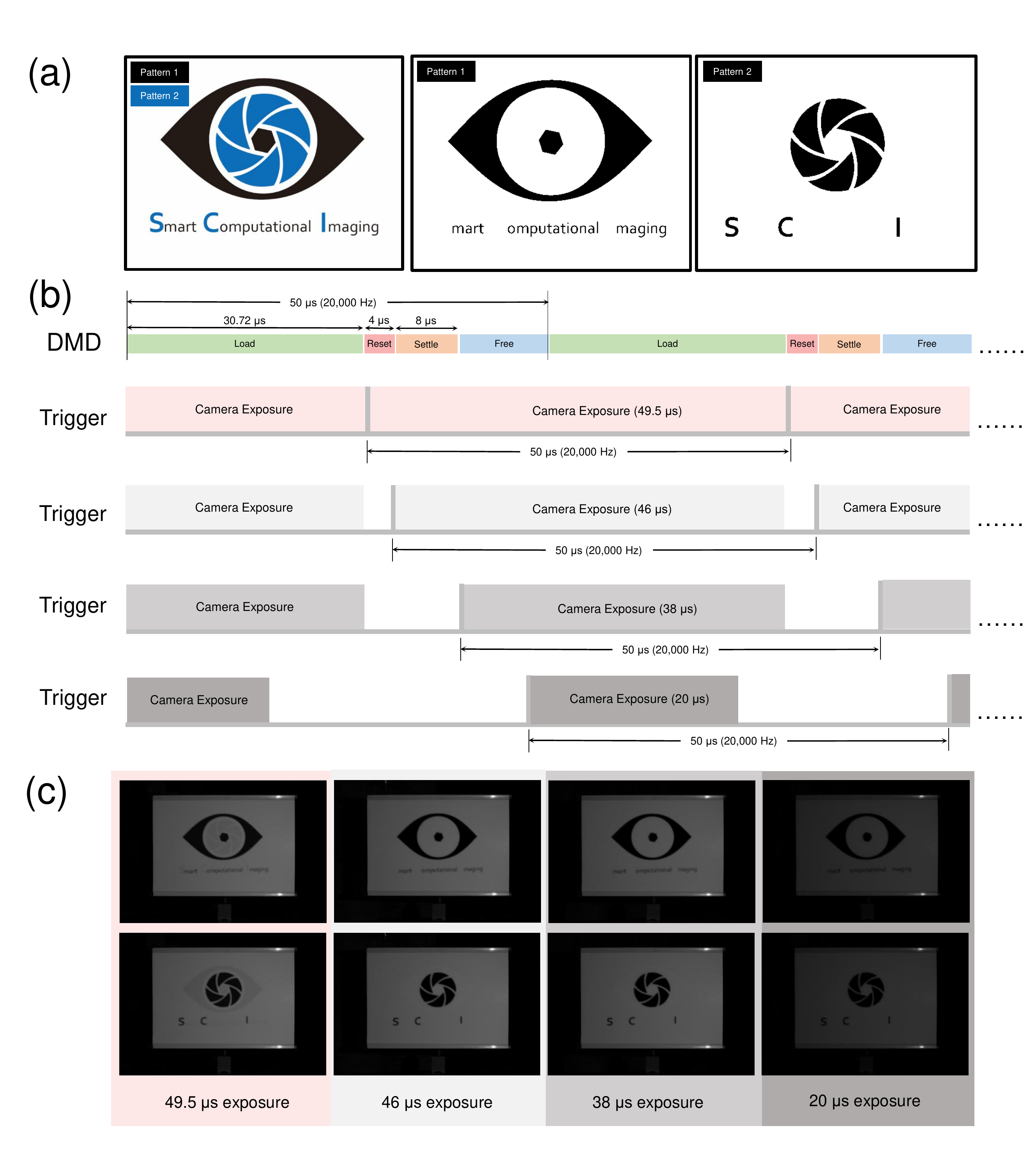}
\caption{Captured images with different exposure and trigger settings. (a) Two complementary patterns which are repeatedly projected by the DLP system at 20,000 Hz. (b) Timing diagrams of different exposure and trigger settings used under the test.  (c) Corresponding images captured by the high-speed camera. Crosstalk can be obviously seen at 49.5 $\mu$s exposure but disappears when the exposure reduces to 46 $\mu$s. Further reducing the exposure offers no visible improvement but leads to light intensity attenuation.}
\label{fig3}
\end{figure}

When both two factors are considered, the total time required to output one pattern on the mirrors is given in Eq. (\ref{2}):
\begin{equation}\label{2}
{\tau  = {\tau _{LD}} + {\tau _{TT}} + {\tau _{ST}} = 30.72 + 4 + 8 = 42.72{\kern 1pt} {\kern 1pt} \mu s}
\end{equation}
so the highest possible rate for binary pattern output is ${1 \mathord{\left/ {\vphantom {1 {42.72{\kern 1pt} {\kern 1pt} \mu s = }}} \right. \kern-\nulldelimiterspace} {42.72{\kern 1pt} {\kern 1pt} \mu s = }}23408$ fps. For the sake of simplicity and security, we update all micro-mirrors with a switching period of 50 $\mu$s, based on the timing shown in Fig. \ref{fig2}. With this timing configuration, we are able to project the binary patterns at 20,000 Hz. However, the synchronization between the DMD operation and the high-speed camera needs to be carefully designed in order to maximize the exposure time while avoid potential cross-talk due to the mirror transition. Thus, we program the trigger output from the DMD to the camera according to the waveform shown in the second row of Fig. \ref{fig2} and set the exposure time of the camera at 46 $\mu$s to skip the mirror transition stage. Further reducing the exposure time does not provide visible improvement but leads to light intensity attenuation, as clearly illustrated in Fig. \ref{fig3}. This is due to the fact that the intensity fluctuation during the 8 $\mu$s mirror settling time can be effectively averaged out during camera exposure. It should also be noted that the trigger signal output from the DLP board is 2.5 V logic level only while the trigger level required for the camera is 5 V. To fix this problem, we build an additional voltage translation circuit (SN74ALVC164245, Texas Instruments) to take the output trigger of the DMD at 2.5 V, condition it, and level convert to the required 5 V for the camera. In  \textbf{Supplementary Video 1}, it is further demonstrated that the designed pattern sequences (see Section \textbf{Principle of $\mu$FTP} for details) can be repeatedly projected onto a dynamic scene (a rotating desk fan) and precisely captured by a synchronized high-speed camera at 20,000 Hz, indicating the synchronization between the DMD and the camera works well.

\subsection*{Projection optics}

\noindent
Since the DLP kit does not include optics, a custom-designed module is attached to the DMD to provide both illumination and projection optics. The projection optics module has two ports, one connecting the DMD and the other connecting the light source. The light source we use is a green LED with an output of 600 lumens (Osram). When the light emitting from the LED enters the projection optics module, it is firstly spatially smoothed by an integration rod to create homogeneous illumination, and then directed onto the DMD by relay optics. The projection lens has a working distance of 800-2,000 mm and an aperture of f/3.8. All lenses components are coated for optimal transmission between 381 and 650 nm. Besides, a copper heat sink, piping, and a fan are added to improve heat dissipation of the whole projection system.

\subsection*{Principle of $\mu$FTP}

\begin{figure}[htb!]
\centering
\includegraphics[width=1.0\linewidth]{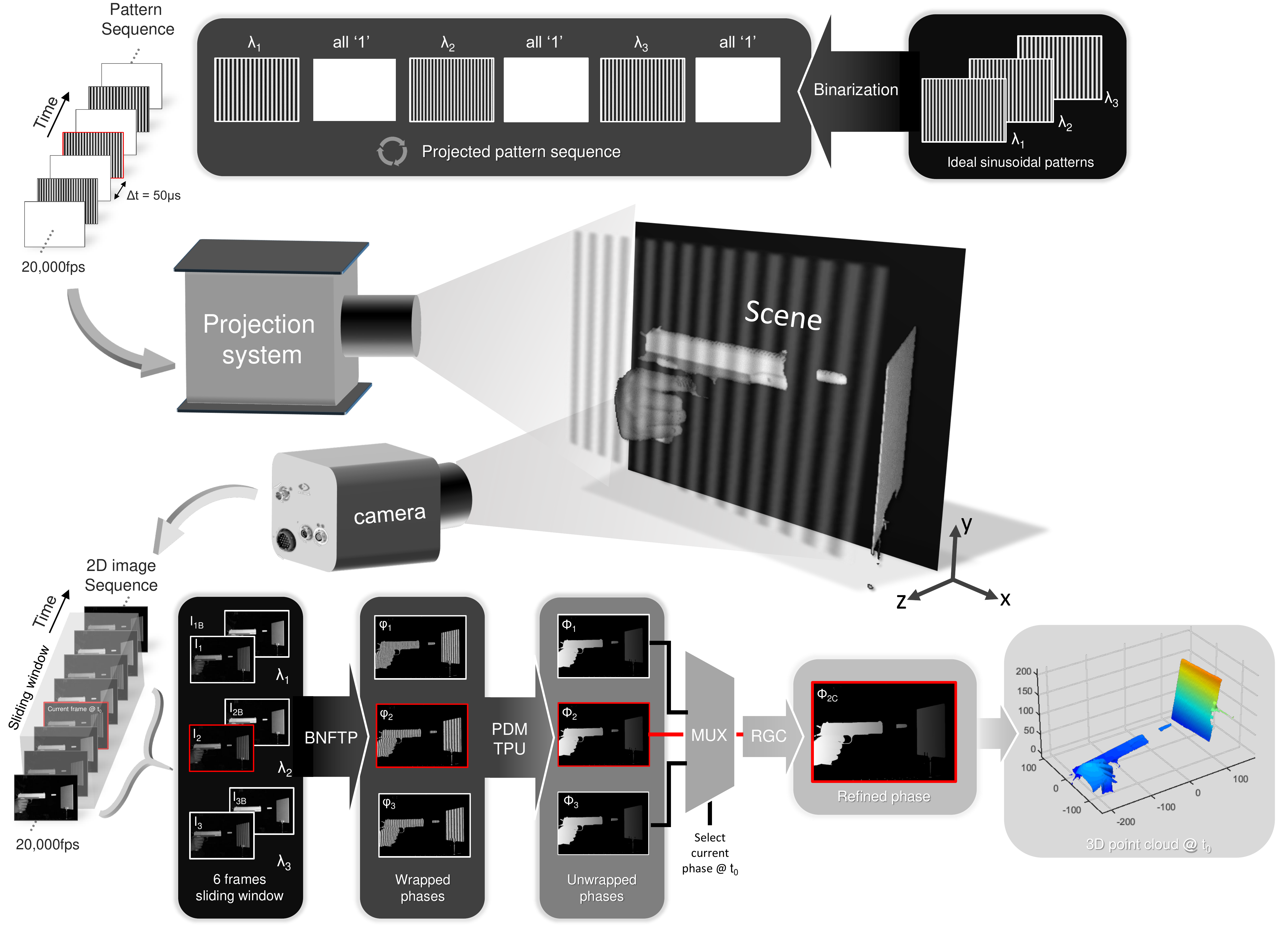}
\caption{Overview of $\mu$FTP framework (frequency number $n = 3$). 3 high-frequency binarized spatial sinusoids with slightly different wavelengths $\lbrace$${\lambda _1}$, ${\lambda _2}$, ${\lambda _3}$$\rbrace$ and 3 white patterns are sequentially projected onto a scene. The camera captures the corresponding synchronized 2D video at 20,000 fps. To reconstruct 3D depth of the scene at a given time ${t_0}$, a 6-frame sliding window centered on the current frame at ${t_0}$ (indicated by the red box) is applied to extracting 3 sinusoid fringe images $\lbrace$${I _1}$, ${I _2}$, ${I _3}$$\rbrace$ and corresponding 3 flat images $\lbrace$${I _{1B}}$, ${I _{2B}}$, ${I _{3B}}$$\rbrace$ with a frame interval of 50 $\mu$s. Three wrapped phase maps with different wavelengths can be recovered by background-normalized FTP (BNFTP) algorithm, and further unwrapped by projection distance minimization temporal phase unwrapping (PDM TPU) algorithm. The phase corresponding to current time point ${t_0}$ (${\Phi _2}$) is selected through multiplexer (MUX) and further refined with a reliability-guided compensation (RGC) algorithm in spatial domain. Finally, the refined phase is used to establish the projector-camera correspondences and reconstruct 3D point cloud through triangulation.}
\label{fig4}
\end{figure}

\noindent
The whole framework of $\mu$FTP is illustrated in Fig. \ref{fig4}, which operates in two stages: acquisition and reconstruction. In the image acquisition stage, $\mu$FTP uses few ($n \ge 2$, e.g., $n = 3$ as illustrated in Fig. \ref{fig1}) high-frequency sinusoidal fringe patterns with slightly different wavelengths (fringe pitches) $\lbrace$${\lambda _1}$, ${\lambda _2}$,…, ${\lambda _n}$$\rbrace$. Between two sinusoids, a ``white" pattern with all mirrors of the DMD in the ``on" state is inserted in the pattern sequence. Thus, there are totally $2n$ patterns that will be rapidly projected onto the measured object surface sequentially at 20,000 fps. To create quasi-sinusoidal gray-scale fringe patterns with the DMD operating in the binary mode, the ideal sinusoids are binarized with Floyd-Steinberg's error diffusion dithering algorithm \cite{73}, and the gray-scale intensity is then reproduced by properly defocusing the projector lens \cite{28,29}. For $\mu$FTP, the fringe wavelength set $\lbrace$${\lambda _1}$, ${\lambda _2}$,…, ${\lambda _n}$$\rbrace$ must meet the following two conditions (see \textbf{Supplementary Information B} for details):

(1) ${\lambda _i}$ should be sufficiently small (frequency is high) as required for successful phase retrieval in conventional FTP based techniques.

(2) The least common multiple (LCM) of the fringe wavelength set should be greater than the total pixel number in the axis wherein the sinusoidal intensity value varies (${LCM\left( {{\lambda _1},{\kern 1pt} {\kern 1pt} {\kern 1pt} {\lambda _2},{\kern 1pt} {\kern 1pt}  \cdots {\kern 1pt} ,{\kern 1pt} {\kern 1pt} {\kern 1pt} {\lambda _n}} \right) \ge W}$) so that the phase ambiguities can be theoretically excluded.

For our DMD projector with a resolution of 1024 $\times$ 768, we find that three wavelengths $\lbrace$${\lambda _1}$, ${\lambda _2}$, ${\lambda _3}$$\rbrace$ = $\lbrace$14, 16, 18$\rbrace$ pixels are sufficient to make a good tradeoff between the fringe contrast and unambiguous phase range (discussion about the wavelength selection is provided in \textbf{Supplementary Information B4}), and they are used throughout the present work.

In the reconstruction stage, the captured pattern images are processed sequentially, with a  $2n$-frame sliding window moving across the whole video sequence. Considering $2n$ frames within a window centered on the current frame (${I_2}$ in Fig. \ref{fig4}) at a given time point (${t_0}$), we have $n$ sinusoid images  $\lbrace$${I_1}$, ${I_2}$, …, ${I_n}$$\rbrace$ and corresponding $n$ white images $\lbrace$${I_{1B}}$, ${I_{2B}}$, …, ${I_{nB}}$$\rbrace$ , as illustrated in Fig. \ref{fig4}. The 3D shape of the measured object at ${t_0}$ is reconstructed from these images based on the 4 steps as follows:

(1) \textbf{Background-normalized Fourier transform profilometry (BNFTP)}: BNFTP is an improved version of conventional FTP \cite{36,38} which is specially designed for high-speed 3D measurement with binary patterns. It uses a sinusoidal fringe image and an additional ``white" image with all “1”s in the projection pattern to recover a high-quality wrapped phase map. The basic theory and implementation of BNFTP is given in \textbf{Supplementary Information A} in details. The advantage of BNFTP over existing FTP-based approaches \cite{36,38,41} lies in the fact that it not only removes the effect of zero-order background but also can handle large surface reflectivity variations (see Figure S1 for experimental comparisons). Considering $n$ sinusoidial images $\lbrace$${I_1}$, ${I_2}$, …, ${I_n}$$\rbrace$ and corresponding $n$ white images $\lbrace$${I_{1B}}$, ${I_{2B}}$, …, ${I_{nB}}$$\rbrace$ within a $2n$-frame sliding window (Fig. \ref{fig4}), we can recover $n$ wrapped phase maps $\left\{ {{\phi _1},{\kern 1pt} {\kern 1pt} {\kern 1pt} {\phi _2},{\kern 1pt} {\kern 1pt}  \cdots {\kern 1pt} ,{\kern 1pt} {\kern 1pt} {\phi _n}} \right\}$ based on BNFTP.

(2) \textbf{Temporal phase unwrapping based on projection distance minimization (PDM)}: The phase maps obtained by Step (1) are wrapped to principle values of the arctangent function, and consequently, phase unwrapping is required to remove the phase ambiguities and correctly extract the object depth. For $\mu$FTP, the phase of the current time point (${t_0}$) is firstly unwrapped temporally by exploiting information from neighboring frames (as shown in Fig. \ref{fig4}, the current phase ${\phi _2}$ is unwrapped with previous and next wrapped phases ${\phi _1}$ and ${\phi _3}$), based on an algorithm so-called PDM (detailed in \textbf{Supplementary Information B}). The basic idea of PDM is to determine the optimum fringe order combination $\lbrace$${k_1}$, ${k_2}$, …, ${k_n}$$\rbrace$ (for each wrapped phase maps $\left\{ {{\phi _1},{\kern 1pt} {\kern 1pt} {\kern 1pt} {\phi _2},{\kern 1pt} {\kern 1pt}  \cdots {\kern 1pt} ,{\kern 1pt} {\kern 1pt} {\phi _n}} \right\}$) so that the corresponding unwrapped phase value combination $\left\{ {{\Phi _1},{\kern 1pt} {\kern 1pt} {\kern 1pt} {\Phi _2},{\kern 1pt} {\kern 1pt}  \cdots {\kern 1pt} ,{\kern 1pt} {\kern 1pt} {\Phi _n}} \right\}$ is ``closest" (in the Euclidean sense) to the following straight line in dimension $n$
\begin{equation}\label{3}
{{\Phi _1}{\lambda _1} = {\Phi _2}{\lambda _2} = ... = {\Phi _n}{\lambda _n}}
\end{equation}
In \textbf{Supplementary Information B2}, we also prove that the phase unwrapping results obtained by PDM algorithm is \emph{optimal in a maximum likelihood sense}. It provides a larger unwrapping range as well as better noise robustness than classic multi-wavelength TPU approaches (e.g. heterodyne approach, as demonstrated in \textbf{Supplementary Information B3 and B4}). Furthermore, the minimum projection distance in PDM also provides an inherent metric to quantitatively evaluate the unwrapping reliability for each pixel, which is used in the following reliability guided compensation algorithm [Step (3)]. It should be also noted that, if the approximate depth range of the measured scene can be estimated, geometric constraint \cite{58,61,62,63} can be applied to restricting the search range for possible fringe orders and ruling out several false candidates beforehand (detailed in \textbf{Supplementary Information D2}). After this step, a group of unwrapped phase maps $\left\{ {{\Phi _1},{\kern 1pt} {\kern 1pt} {\kern 1pt} {\Phi _2},{\kern 1pt} {\kern 1pt}  \cdots {\kern 1pt} ,{\kern 1pt} {\kern 1pt} {\Phi _n}} \right\}$ can be obtained, and only the phase map corresponding the current time point ${t_0}$ (${\Phi _2}$ shown in Fig. \ref{fig4}) will be further processed by the following steps.

(3) \textbf{Reliability guided compensation (RGC) of fringe order error}: Although the initial unwrapped phase map obtained by Step (2) (${\Phi _2}$ shown in Fig. \ref{fig4}) encodes depth information of the measured scene at ${t_0}$ with a temporal resolution of 100 $\mu$s (remember that in Step (1) the principal value of ${\Phi _2}$ is recovered only from 2 patterns), fringe order errors are inevitable especially around dark regions (lower fringe quality) and object edges (higher motion sensitivity). We propose an approach called RGC for identifying and compensating those fringe order errors by exploiting additional information in the spatial domain (detailed in \textbf{Supplementary Information C}). Our key observation is that the fringe order errors are usually isolated (at least less concentrated than the correct phase values) delta-spike artifacts with a phase error of integral multiples of $2\pi$. Inspired by quality guided (spatial) phase unwrapping approaches \cite{74,75,76}, we first gather neighboring pixels within a continuous region of the phase map into groups. Then the isolated pixels or pixels falling into small groups are considered as fringe order errors, and their phase values will be corrected sequentially according to an order ranked by a predefined reliability function (we adopt the minimum projection distance in PDM algorithm as the reliability function, and larger distance means lower reliability). After RGC compensation, we can obtain the refined absolute phase map that is free from fringe order errors (${\Phi _{2C}}$ shown in Fig. \ref{fig4})

(4) \textbf{Mapping from phase to 3D coordinates}: The final step of $\mu$FTP reconstruction is to establish the projector-camera pixel correspondences based on the refined absolute phase map (${\Phi _{2C}}$ shown in Fig. \ref{fig4}) and to reconstruct the 3D coordinates of the object surfaces at time ${t_0}$ based on the calibration parameters of the projector and the camera. In this process, the effect of projection-imaging distortion of lenses is explicitly considered and effectively corrected by an iterative scheme based on lookup table (LUT) implementation (described in \textbf{Supplementary Information D}). For more details about the implementation of $\mu$FTP, one can refer to the \textbf{MATLAB source code} available on our website \cite{77}.

Compared with existing techniques, $\mu$FTP has the prominent advantage of recovering an accurate, unambiguous, and dense 3D point cloud with only two projected patterns. Furthermore, since each phase is encoded within a single high-frequency sinusoidal pattern, $\mu$FTP can be considered “single-shot" in essence, allowing for motion-artifact-free 3D reconstruction with fine temporal resolution (2 patterns per 3D reconstruction, corresponding to 10,000 3D fps). It is also worth mentioning that since $\mu$FTP is based on sliding-window reconstruction, any newly added image can be combined with its preceding $2n - 1$ images to produce a new 3D result. Although in this case the consecutive two 3D reconstructions may use the same high-frequency fringes for phase evaluation, one can achieve a pseudo frame rate of 20,000 3D fps, which is same as the projector and camera speed.

\section*{Results}

\subsection*{Quantitative analysis of 3D reconstruction accuracy}

\begin{figure}[htb!]
\centering
\includegraphics[width=1.0\linewidth]{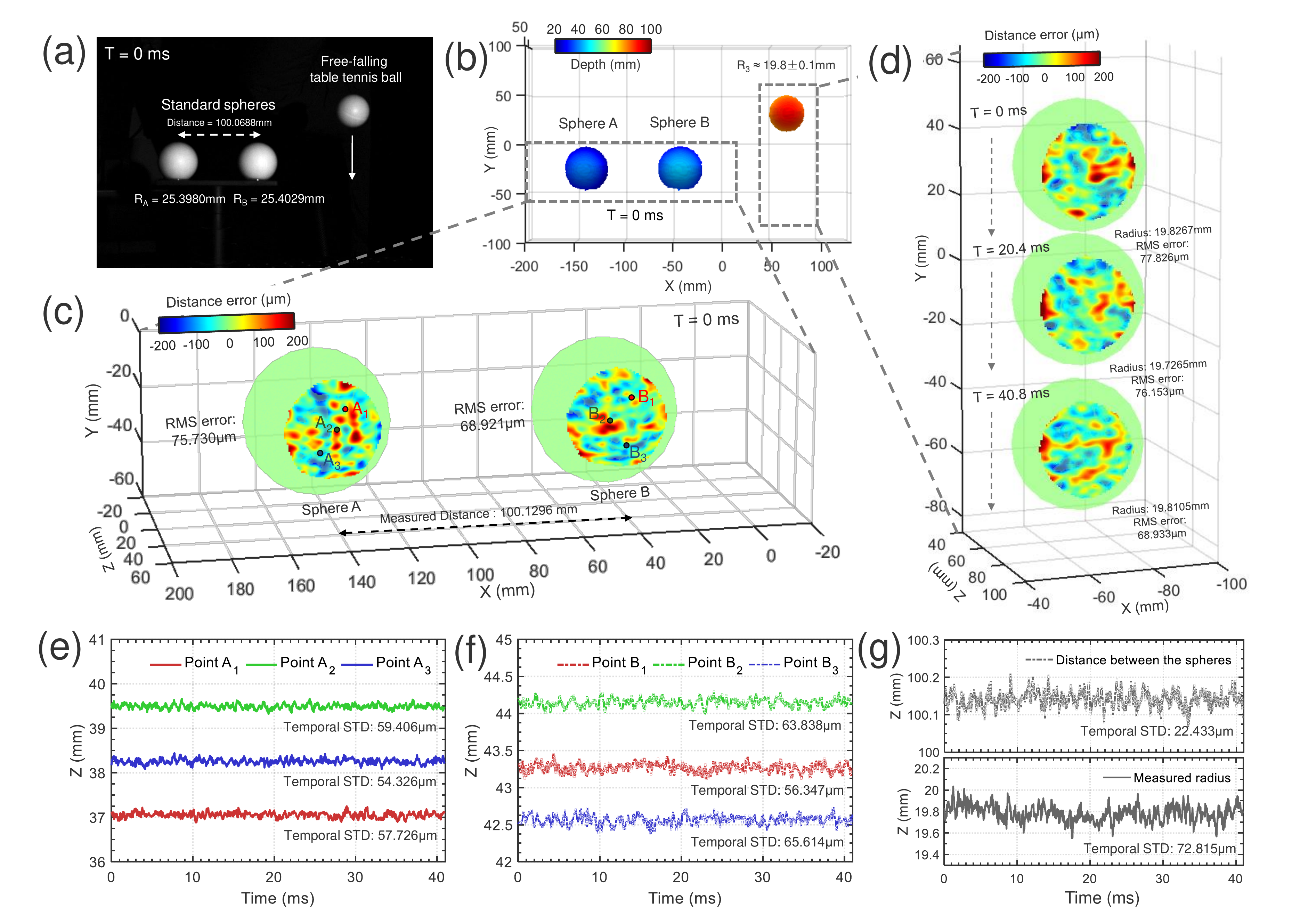}
\caption{3D reconstruction accuracy of $\mu$FTP measurement. (a) Test scene consisting of two standard ceramic spheres and a free-falling table tennis ball at T = 0 ms. (b) Corresponding color-coded 3D reconstruction. (c) The error distribution of the measured standard spheres at T = 0ms. (d) The deviation map (the difference between the measured points and the fitted spheres) of the free-falling table tennis ball at three different time points (T = 0 ms, 20.4 ms, 40.8 ms). (e) Repeatability of 820 measurements over a 41 ms period of 3 points (A$_1$, A$_2$, and A$_3$) on the standard Sphere A. (f) Repeatability of 820 measurements of 3 points (B$_1$, B$_2$, and B$_3$) on the standard Sphere B. (g) Repeatability of 820 measurements of the center-to-center distance between the two spheres as well as the radius of the free-falling table tennis ball.}
\label{fig5}
\end{figure}

\noindent
To quantitatively determine the accuracy of $\mu$FTP measurement, we conduct an experiment on a test scene consisting of two standard ceramic spheres and a free-falling table tennis ball, as shown in Fig. \ref{fig5}(a). The measured objects are put in the measurement volume, which is approximately 400 mm $\times$ 275 mm $\times$ 400 mm, established by using a calibration panel (see \textbf{Supplementary Information D3} for details). According to the calibration based on a coordinate measurement machine (CMM), the radii of the two standard spheres are 25.3980 mm and 25.4029 mm, respectively, and their center-to-center distance is 100.0688 mm. These two spheres with accurate calibrated dimensions are used to quantify the measurement accuracy and repeatability of $\mu$FTP system. The table tennis ball is used to test the performance of the system for measuring moving object, whose radius is about 19.8 $\pm$ 0.1 mm (dimension uncalibrated, radius is simply measured by a vernier caliper). Figure \ref{fig5}(b) shows the color-coded 3D reconstruction by $\mu$FTP at T = 0 ms and Fig. \ref{fig5}(c) shows the corresponding error distribution of the two measured standard spheres. The accuracy is distinguished by fitting the standard sphere to the point cloud representing the spherical surface and calculating the difference between the measured points and the fitted sphere. As shown in Fig. \ref{fig5}(c), the root mean square (RMS) errors corresponding to the two standard spheres are 75.730 $\mu$m and 68.921 $\mu$m, respectively. The measured center-to-center distance between the two spheres is 100.1296 mm. In Fig. \ref{fig5}(d), we further show the deviation maps of the free-falling table tennis ball at three different time points (T = 0 ms, 20.4 ms, 40.8 ms). Note that since the dimension of the table tennis ball is uncalibrated, we determine the best spherical fit on the 3D point cloud to estimate the spherical diameter, and the deviation maps shown in Fig. \ref{fig5}(d) are the difference between the measured points and the fitted spheres. These results indicate that the measurement accuracy of $\mu$FTP is better than 80 $\mu$m for both static and dynamic objects. Repeatability of the $\mu$FTP measurement is also analyzed by performing 820 measurements over a 41 ms period (at a pseudo frame rate of 20,000 fps) for 3 different points on respective standard sphere [A$_1$ $\sim$ A$_3$ and B$_1$ $\sim$ B$_3$, as labelled in Fig. \ref{fig5}(c)], the center-to-center distance between the two standard spheres, and the radius of the free-falling table tennis ball, as shown in Figs. \ref{fig5}(e)-\ref{fig5}(g), respectively. The temporal movie of color-coded 3D reconstruction of the test scene and the corresponding error analysis over the 41 ms period is further provided in \textbf{Supplementary Video 2}. One can clearly observe the excellent repeatability of $\mu$FTP system: the center-to-center distance measurement exhibited a very low temporal standard deviation (STD) of 22.433 $\mu$m; the temporal STD at one given point on the standard sphere is typically around 60 $\mu$m; the radius of the free-falling table tennis ball has a slightly higher temporal STD (72.815 $\mu$m) due to the object motion. These results show that the presented system achieves a measurement accuracy better than 80 $\mu$m and a temporal STD below 75 $\mu$m according to the measurement volume of 400 mm $\times$ 275 mm $\times$ 400 mm at 10,000 3D fps.

\subsection*{Vibrating cantilevers}

\begin{figure}[htb!]
\centering
\includegraphics[width=0.9\linewidth]{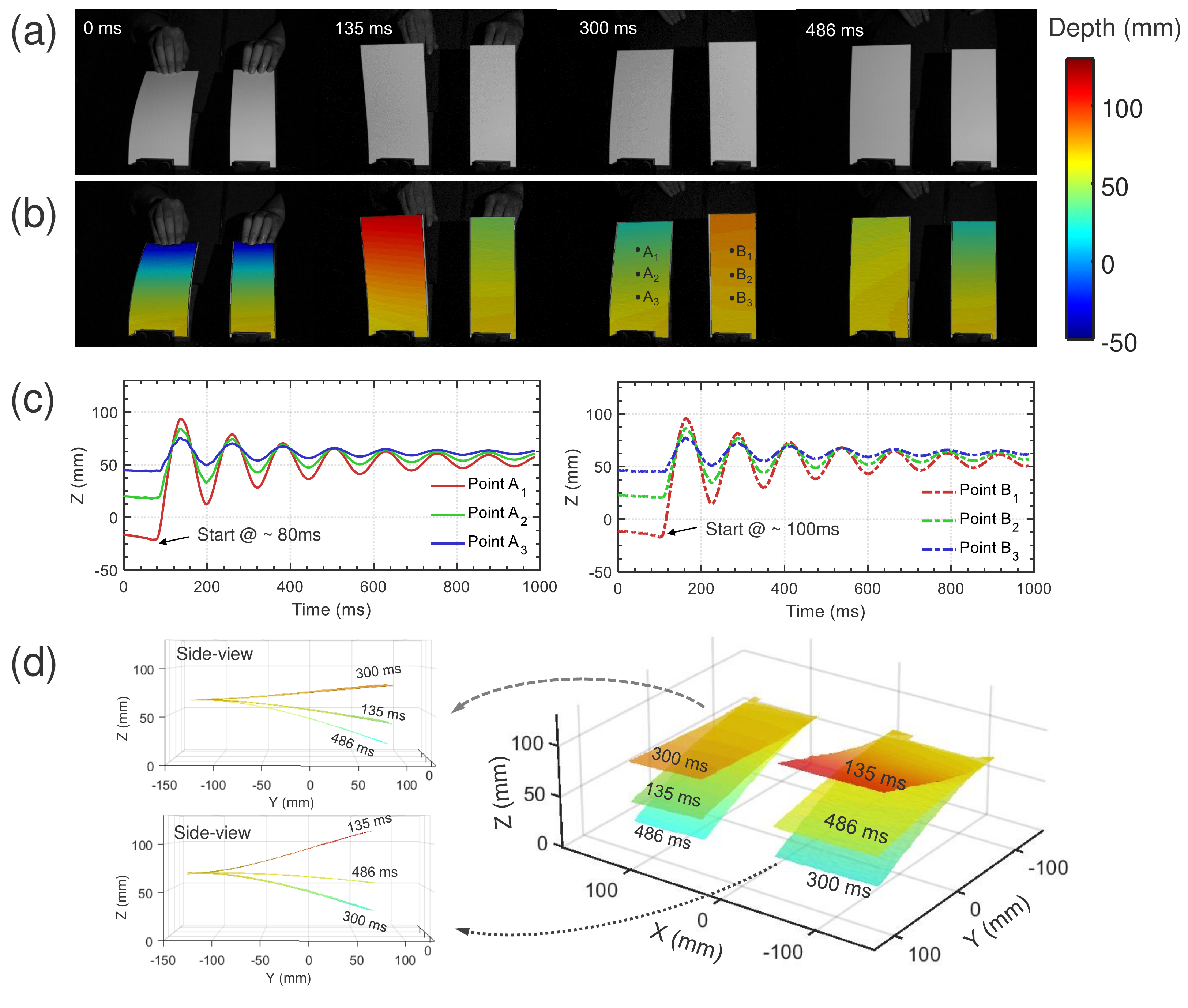}
\caption{Measurement of two vibrating cantilever beams. (a) Representative camera images at different time points. (b) Corresponding color-coded 3D reconstructions. (c) Displacement of 3 points on each cantilever [A$_1$ $\sim$ A$_3$ and B$_1$ $\sim$ B$_3$, as labelled in (b)] as a function of time. (d) The 3D geometric field of the two cantilevers at three different time points. The two insets show the side-views (y-z plane) of the respective cantilever.}
\label{fig6}
\end{figure}

\noindent
In this subsection, vibrating cantilever beams are used to validate $\mu$FTP’s fast 3D imaging capability and compare its performance with the state-of-the-arts. Firstly, two fixed-end homogeneous plastic cantilever beams are imaged. The dimensions of the two cantilever beams are 215 mm $\times$ 110 mm $\times$ 1.5 mm [for the larger one shown on the left of Fig. \ref{fig6}(a)] and 215 mm $\times$ 80 mm $\times$ 1.5 mm [for the smaller one shown on the right of Fig. \ref{fig6}(a)]. One end of each cantilever is clamped on the optical table while the other end is excited by human hand. Figures \ref{fig6}(a) and \ref{fig6}(b) show representative camera images (white pattern) and corresponding color-coded 3D reconstructions by $\mu$FTP at different time points. Initially, the two cantilevers are heavily bent with manual pressure, with their ends aligned at about the same depth. When hands release, the stored elastic potential energy converts to kinetic energy to enable vibrations. To study the vibrating process quantitatively, the out-of-plane (z) displacement of 3 points on each cantilever [A$_1$ $\sim$ A$_3$ and B$_1$ $\sim$ B$_3$, as labelled in Fig. \ref{fig6}(b)] are plotted as a function of time in Fig. \ref{fig6}(c). The plots show that the largest vibration amplitudes occur at points A$_1$ and B$_1$, since they are closer to the free ends of the cantilevers. Their vibration amplitudes gradually reduced from about 50 mm to 10 mm (for point A$_1$ and B$_1$) with a frequency of $\sim$ 8 Hz. Besides, there is about 20 ms time difference between the two hand releases, making the vibrations of the two cantilever out-of-phase. In Fig. \ref{fig6}(d), we show the reconstructed 3D shapes of the two cantilevers at three different time points, with the two insets showing the side-views (y-z plane) of the respective cantilever. The movie of color-coded 3D rendering of the two cantilevers’ surfaces as well as the corresponding side-views is provided in \textbf{Supplementary Video 3}. These results verify that the proposed $\mu$FTP enables high-speed 3D reconstruction of object vibration and provides high-accuracy quantitative evaluation of any characteristic points on the object surface.

\begin{figure}[htb!]
\centering
\includegraphics[width=1.0\linewidth]{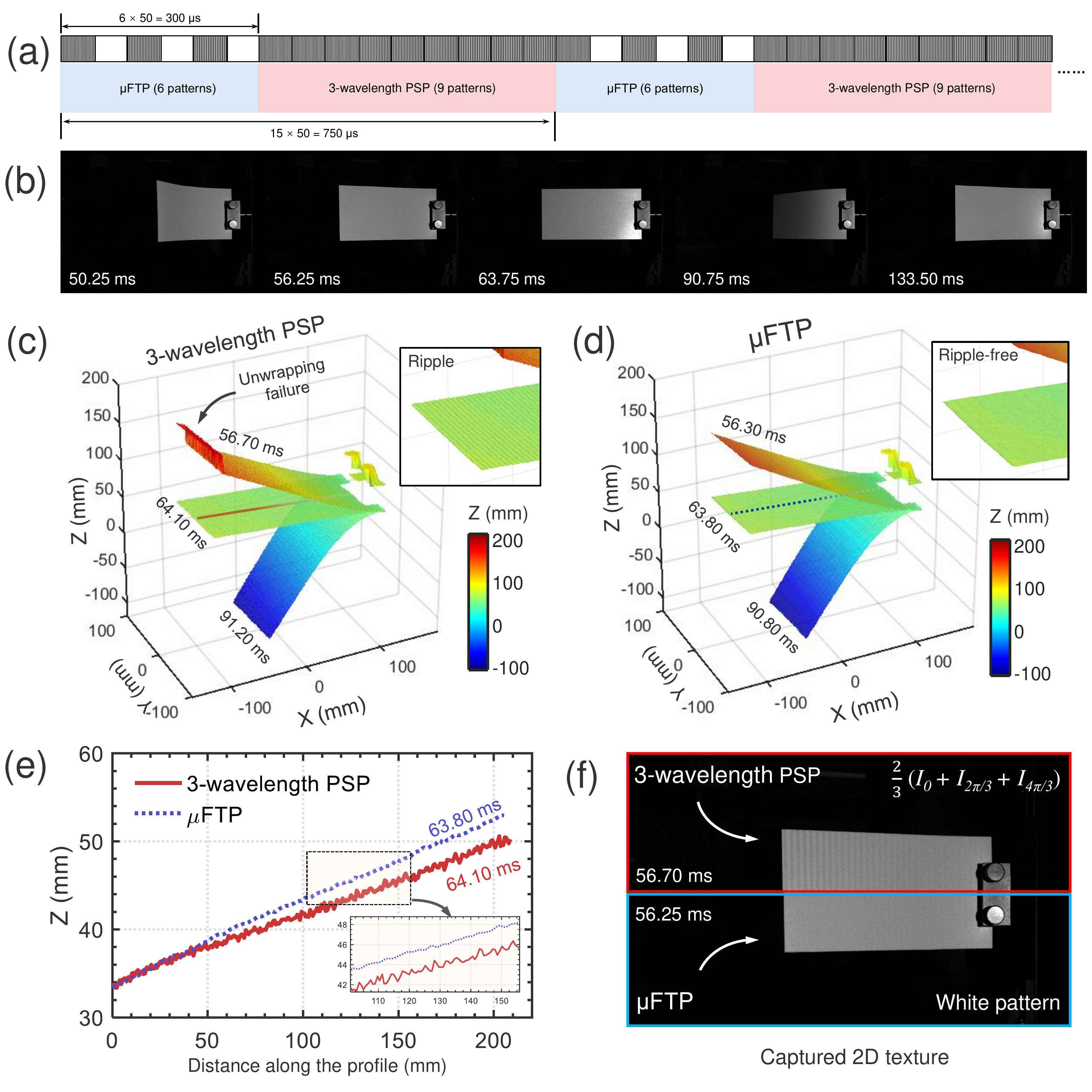}
\caption{Comparison of $\mu$FTP with three-wavelength PSP. (a) Projected pattern sequence (6 patterns for $\mu$FTP and 9 pattern for 3-wavelength PSP). (b) Representative camera images (white pattern from $\mu$FTP) at different time points. (c) 3D reconstructions of the cantilever surface by 3-wavelength PSP at three different time points. (d) 3D reconstructions of the cantilever surface by $\mu$FTP at three different time points.  (e) Line profiles along the central lines of cantilever surface [corresponding to the dashed lines in (c) and (d)]. (f) Comparison of 2D texture obtained by three-wavelength PSP and $\mu$FTP.}
\label{fig7}
\end{figure}

Next, we compare our $\mu$FTP with the well-known three-wavelength phase shifting profilometry (PSP) \cite{47,53} by using only the larger cantilever beam. One end of the cantilever is vertically fixed on the optical table, and the free end is heavily excited to create large amplitude vibration. In order to achieve a fair comparison, we project the required 15 patterns for both $\mu$FTP (6 patterns) and three-wavelength PSP (9 patterns) sequentially onto the same scene, according to the pattern sequence illustrated in Fig. \ref{fig7}(a). Both algorithms use the same fringe wavelengths $\lbrace$${\lambda _1}$, ${\lambda _2}$, ${\lambda _3}$$\rbrace$ = $\lbrace$14, 16, 18$\rbrace$ pixels, and the retrieved phases are processed with the same algorithms (PDM unwrapping and RGC), respectively. The final 3D point clouds are both reconstructed from the respective unwrapped phases corresponding to ${\lambda _2}$. Since the two groups of patterns are sequentially projected onto the moving surface, there exists a 400 $\mu$s time difference between the two 3D reconstructions from these two methods. Figure \ref{fig7}(b) shows representative camera images (white patterns from $\mu$FTP) at different time points. In Figs. \ref{fig7}(c) and \ref{fig7}(d), we compare the reconstructed 3D surfaces of the cantilever at three different time points. As clearly highlighted in the zoom area and the line profile [Fig. \ref{fig7}(e)], apparent ripples can be observed in the results of 3-wavelength PSP, which also cause unwrapping errors around the end of cantilever (where the vibration amplitude is high). Whereas the $\mu$FTP produces decent 3D reconstruction without notable motion ripples (see also the \textbf{Supplementary Video 4} for a comparative movie). It should be pointed out that the flat image in $\mu$FTP is inherently a normal 2D image that can be used for texture mapping. However, as shown in Fig. \ref{fig7}(f), the fringe images have to be averaged to create a texture for PSP, which is also easily distorted by the object motion (see also the \textbf{Supplementary Video 5} for a comparative movie of the 3D cantilever surfaces with texture mapping). These results verify that $\mu$FTP is completely immune to motion ripples \cite{46,50}, leading to distortion-free 3D reconstruction along with high-quality 2D texture even though the motion is fast and the out-of-plane displacement is large.

\begin{figure}[htb!]
\centering
\includegraphics[width=1.0\linewidth]{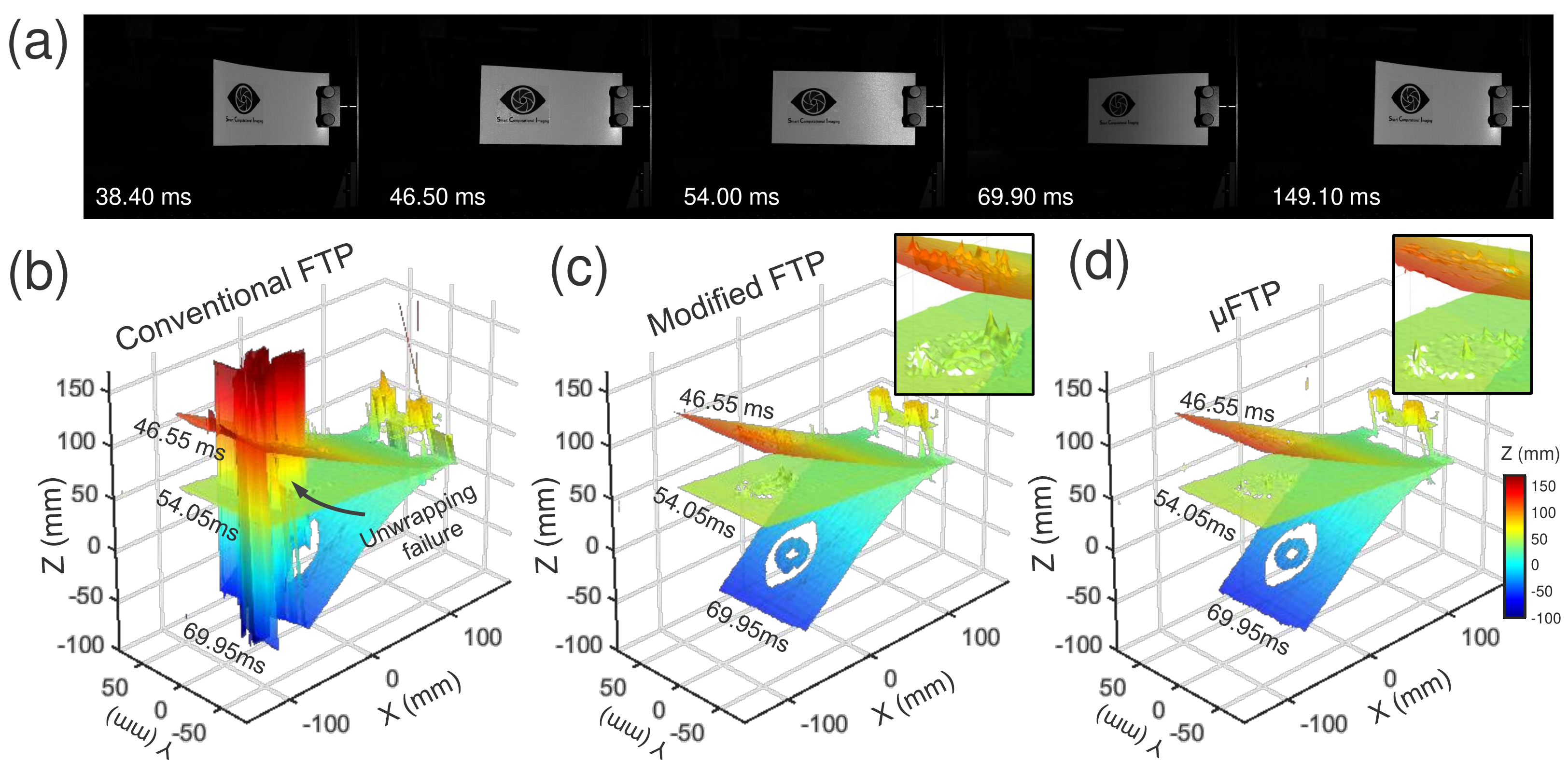}
\caption{Comparison of conventional FTP, modified FTP (background subtraction), and $\mu$FTP (background subtraction and normalization). (a) Representative camera images at different time points. (b) 3D reconstructions of the cantilever surface with conventional FTP at three different time points. (c) 3D reconstructions with modified FTP. (d) 3D reconstructions with $\mu$FTP.}
\label{fig8}
\end{figure}

Finally, we compare our $\mu$FTP with conventional FTP \cite{36,38} and modified FTP \cite{41}. The measured scene is similar with the previous demonstration but an additional sticker (our lab logo) is pasted on the cantilever surface to create large reflectivity variations [Fig. \ref{fig8}(a)]. All three algorithms use the same raw image data (the white images are not used in conventional FTP), and the retrieved phases are processed with the same algorithms (PDM unwrapping and RGC). The reconstructed 3D surfaces of the cantilever at three different time points by using the three methods are shown in Figs. \ref{fig8}(b)-\ref{fig8}(d), respectively. As can be seen, the results generated by conventional FTP suffer from obvious artifacts and unwrapping errors around sharp edges and dark regions. Modified FTP eliminates the unwrapping error and reduces the artifacts. But there remain severe fluctuations (highlighted in the zoom area), which is mainly caused by large reflectivity variations. In contrast, $\mu$FTP produces a much smoother reconstruction without notable artifacts. These results suggest that $\mu$FTP notably improves the performance of state-of-the-arts in terms of accuracy and robustness.

\subsection*{Rotating fan blades}
\begin{figure}[htb!]
\centering
\includegraphics[width=1.0\linewidth]{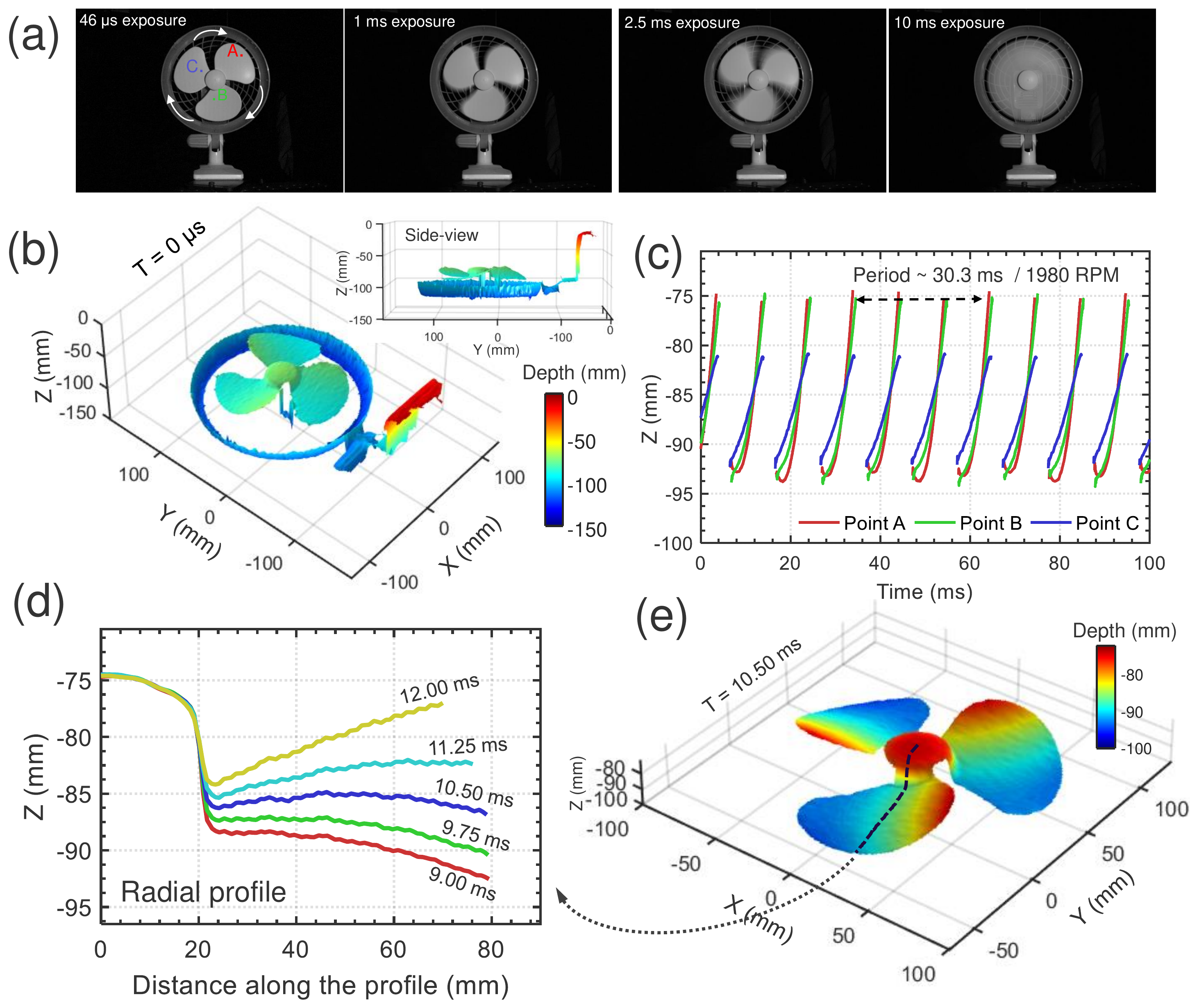}
\caption{Measurement of rotating blades. (a) Images captured by the camera at different exposure time (46 $\mu$s, 1 ms, 2.5 ms, and 10 ms). (b) The reconstructed 3D shape of the whole fan at the start of the observation time (T = 0 ms), the inset shows the side-view (y-z plane) of the 3D reconstruction. (c) Displacement in the z direction at 3 chosen point locations [A, B, and C, shown in (a)] as a function of time over a 100 ms period. (d) 5 line profiles drawn along the radial direction [corresponding to the dashed line in (e)] at time intervals of 0.75 ms. (e) The color-coded 3D rendering of the fan’s surface at T = 10.5 ms.}
\label{fig9}
\end{figure}

\noindent
The next test object is a commercially available desk fan with 3 blades made of $\sim$ 1.6 mm thick plastic. The fan is fixed on the optical table with its front protect shell removed so that the fan blades can be directly exposed to the measurement system. As shown in Fig. \ref{fig9}(a), though the fan is rotating at its highest possible speed, the 46 $\mu$s exposure time of $\mu$FTP system is short enough to “freeze” the high-speed motion and record a clear image of the whirling fan blades. To illustrate the rotating speed of the fan blades more intuitively, we increase the camera exposure from 46 $\mu$s to 1 ms, 2.5 ms, and 10 ms. The blade edges become increasingly blurred. And finally at 10 ms exposure, we are unable to identify the fan blades since the motion blur makes them to appear as one streak. Despite of this challenging rotation speed, the 3D shape of the whole fan, including the base, center hub, side cover, and three blades are well reconstructed by $\mu$FTP, as shown in Fig. \ref{fig9}(b). One point on each blade is chosen to demonstrate the cyclic displacements of the fan blades [points A, B, and C, shown in Fig. \ref{fig9}(a)]. Displacement in the z (out-of-plane) direction at the chosen point locations are plotted as a function of time over a 100 ms period, as shown in Fig. \ref{fig9}(c). The plot shows that the fan has a rotation period of approximately 30.3 ms, corresponding to a speed of 1,980 rotations per minute (rpm). The plot also demonstrates a good repeatability of the $\mu$FTP measurement. Besides, by applying a proper threshold on the captured white pattern, a binary mask can be generated to extract the moving fan blades region of interest from the static background. The color-coded 3D rendering of the fan’s surface at one point in time (10.5 ms) is illustrated in Fig. \ref{fig9}(e). Figure \ref{fig9}(d) further gives 5 line profiles drawn along the radial direction out from the center hub [corresponding to the dashed line in Fig. \ref{fig9}(e)] at time intervals of 0.75 ms. Within the short 3 ms, the fan blade quickly sweeps through the radical profile for about 1/10 of a revolution, resulting in a maximum variation over 15 mm in the z direction. The results also demonstrate that the length of fan blade is $\sim$ 80 mm in the radial direction out from the center hub which has a radius of $\sim$ 20 mm. The corresponding 3D movie is further provided in \textbf{Supplementary Video 6}. It is important to mention that unlike the previous study based on stroboscopic structured illumination \cite{78}, here we truly recorded the entire process of fan rotation without any stroboscopic time gap between two successive 3D frames.

\subsection*{Bullet fired from a toy gun}

\begin{figure}[htb!]
\centering
\includegraphics[width=1.0\linewidth]{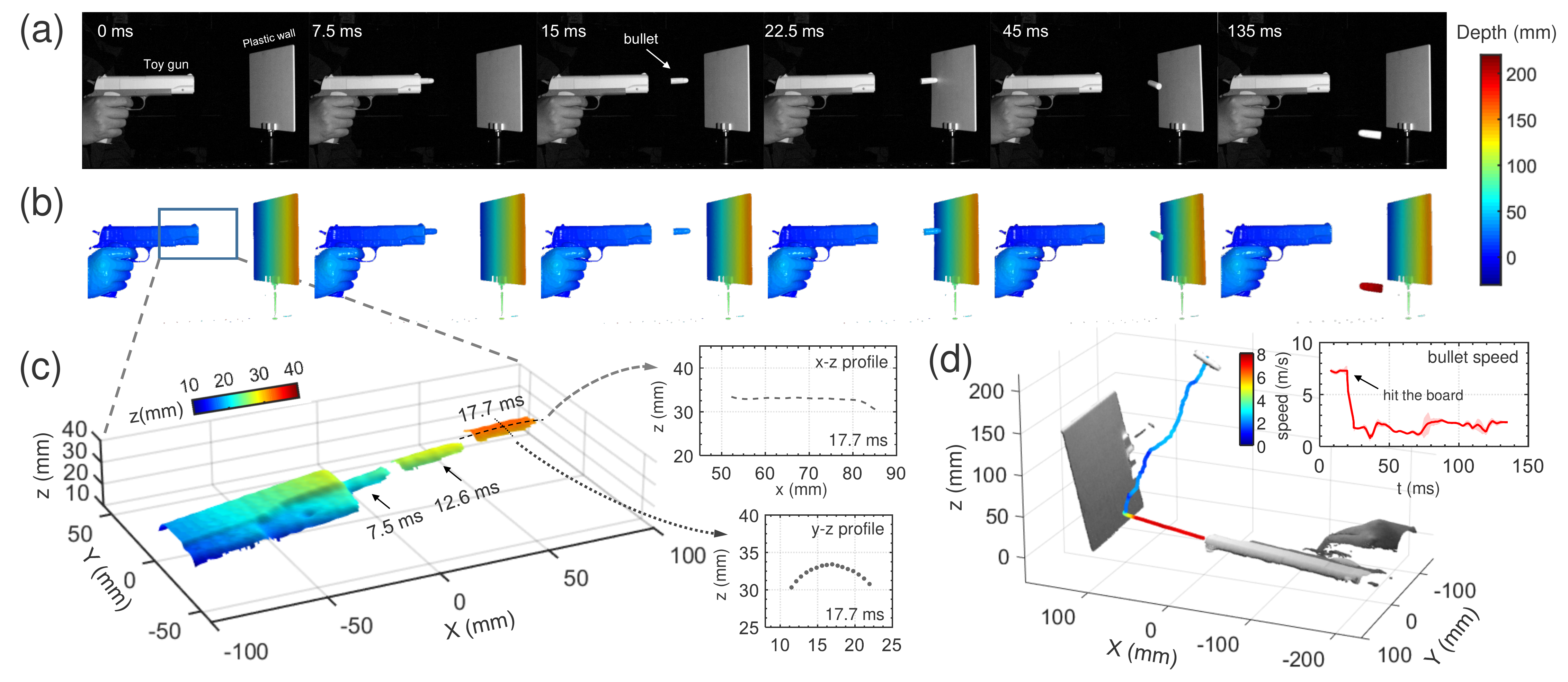}
\caption{3D measurement and tracking a bullet fired from a toy gun. (a) Representative camera images at different time points. (b) Corresponding color-coded 3D reconstructions. (c) 3D reconstruction of the muzzle region [corresponding to the boxed region shown in (b)] as well as the bullet at three different points of time over the course of flight (7.5 ms, 12.6 ms, and 17.7ms). The insets show the horizontal (x-z) and vertical (y-z) profiles crossing the body center of the flying bullet at 17.7 ms. (d) The 3D point cloud of the scene at the last moment (135 ms), with the colored line showing the 130 ms long bullet trajectory. The inset plots the bullet velocity as a function of time.}
\label{fig10}
\end{figure}

\noindent
Next, we apply $\mu$FTP to image one-time transient event: a bullet fired from a toy gun and then rebounded from a plaster wall. Figures \ref{fig10}(a)-\ref{fig10}(b) show representative camera images (white pattern) and corresponding color-coded 3D reconstructions at different time points. T = 0 ms is the start of the observation time, and the bullet begins to occur in the vicinity of the gun muzzle at about T = 7.5 ms. After travelling in free-flight for about 15 ms, the bullet hits the plaster and rebounds towards the camera. In Fig. \ref{fig10}(c), we show the 3D reconstruction of the muzzle region [corresponding to the boxed region in Fig. \ref{fig10}(b)] as well as the bullet at three different points of time (7.5 ms, 12.6 ms, and 17.7 ms). The two insets further provide the horizontal (x-z) and vertical (y-z) profiles crossing the body center of the bullet at 17.7 ms, which indicate that the bullet has a length of $\sim$ 35.5 mm and a diameter of $\sim$ 11.8 mm. Besides, the 3D data can be used to quantitatively analyze the process with regards to the ballistic trajectory and velocity. By tracing the center of the bullet body, we can obtain the bullet trajectory in 3D space. The instantaneous speed of the bullet can then be estimated by taking the derivative of the position function with respect to time. The calculated muzzle velocity (velocity of the bullet when it leaves the barrel) is around 7.3 m/s. At that speed, the bullet moves about one pixel between each camera frames. Since the phase information is encoded in single fringe pattern in $\mu$FTP, the frame-by-frame motion does not introduce any visible artifacts. In Fig. 10(d), we further show the 3D point cloud of the scene at the last moment (T = 135 ms), with the colored line showing the 130 ms long bullet trajectory (the bullet velocity is encoded by the line color, see \textbf{Supplementary Video 7} for time evolution of the trajectory). The inset of Fig. \ref{fig10}(d) provides the bullet velocity as a function of time, indicating that the bullet speed keeps almost constant when travelling in free-flight, and suddenly reduces to about 2 m/s during the collision. The fluctuation of speed after collision mainly results from the rolling over of the bullet, which also increases the estimation uncertainty [red shaded region in Fig. \ref{fig10}(d)] due to the difficulties in accurate tracking of the bullet body center. A more detailed illustration of the transient event is provided in \textbf{Supplementary Video 8}, which is a slow-motion 3D movie containing 2,700 3D frames with a frame interval of 50 $\mu$s (corresponding to 20,000 3D fps over an observation period of 135 ms). These experimental results demonstrate the potential applications of $\mu$FTP for tracking 3D trajectory of fast moving object within a wide observation volume.

\subsection*{Balloon’s explosion triggered by a flying dart}

\begin{figure}[htb!]
\centering
\includegraphics[width=1.0\linewidth]{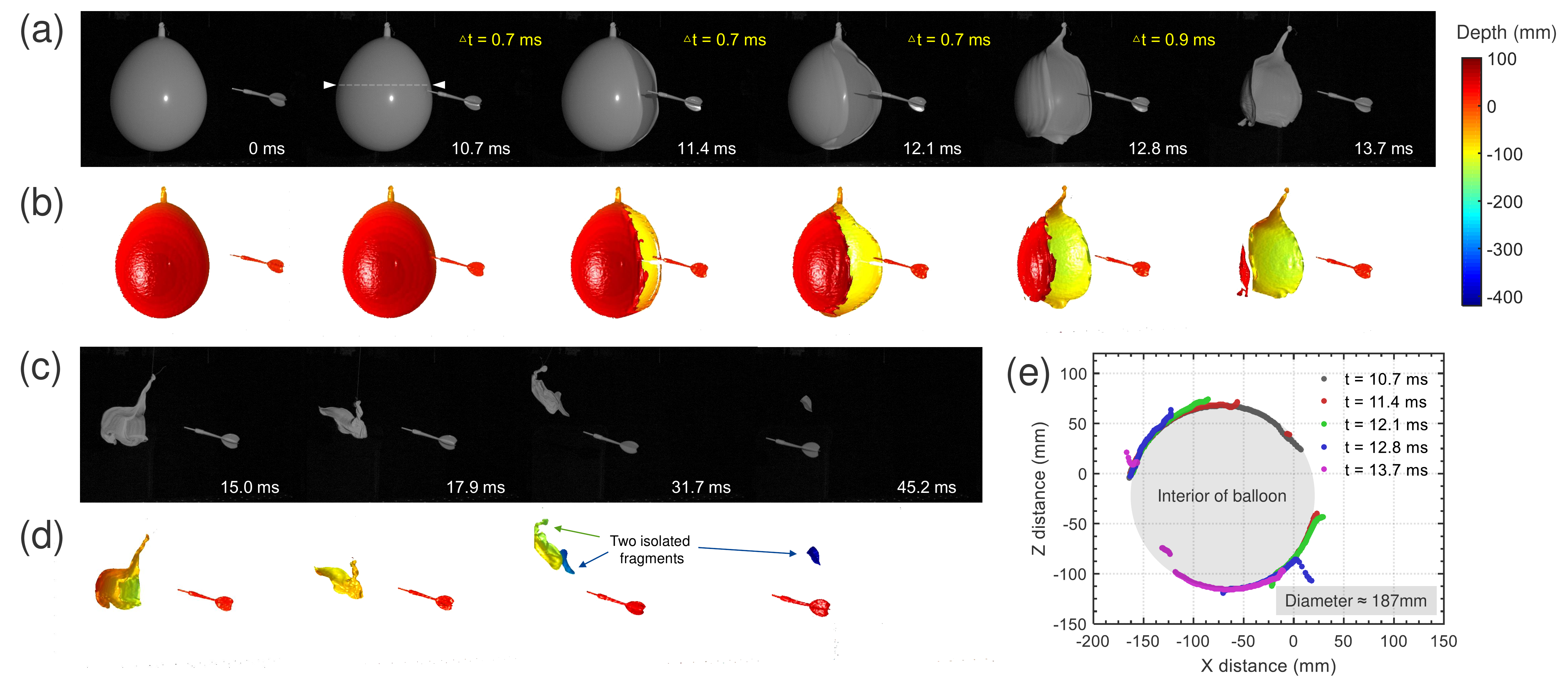}
\caption{Balloon’s explosion triggered by a flying dart. (a) Representative camera images at different time points. (b) Corresponding color-coded 3D reconstructions. (c) and (d) continue to (a) and (b), respectively. (e) Line profiles across the dashed line in (a) at the time points of 10.7 ms, 11.4 ms, 12.1 ms, 12.8 ms and 13.7 ms.}
\label{fig11}
\end{figure}

\noindent
In the last demonstration, the proposed $\mu$FTP is applied to capturing a very-high-frequency event — air balloon bursting punctured by a flying dart. Figures \ref{fig11}(a)-\ref{fig11}(d) show representative 2D camera images (white pattern) and corresponding color-coded 3D reconstructions at different time points. The corresponding movie is provided in \textbf{Supplementary Video 9}. The balloon is suspended in air by threads and keeps still until its surface touchs the tip of the flying dart (T = 10.7 ms). The pricked hole propagates into a radial crack towards to two poles of the balloon, slicing the balloon into a piece of rubber membrane (T = 13.7 ms. The membrane then shrinks into a crumpled form rapidly (T = 15 ms) and finally breaks into 2 fragments (T = 31.7 ms, indicated by the yellow and blue arrows). The whole process lasts about 47 ms, while the key event (balloon blowing up) takes place only within about 4 ms (note that the frame intervals from the second to sixth 3D snapshots in Fig. \ref{fig11}(b) are less than 1 ms). The 3D data can be used to quantitatively analyze the explosion process. Figure \ref{fig11}(e) shows 5 line profiles across the dashed line shown in Fig. \ref{fig11}(a), corresponding to the time points of 10.7 ms, 11.4 ms, 12.1 ms, 12.8 ms and 13.7 ms. When the balloon is intact, only the top surface can be imaged (shown in gray). With the crack expanding and propagating, the bottom (inner) surface of the balloon is revealed. It is interesting that, except for the eversion around the crack boundaries, the main balloon surface still demonstrates a good axi-symmetry during explosion (3 ms), characterized by a longitudinal diameter of $\sim$ 187 mm [shaded region in Fig. \ref{fig11}(d)]. In \textbf{Supplementary Video 9}, we can see that the depth information of the sudden explosion within time spans on the order of tens of microseconds is fully recovered by $\mu$FTP. The 3D results show good image quality without depth ambiguities (note in Figs. \ref{fig11}(c) and \ref{fig11}(d), the two ``overlapping" fragments in 2D camera images are actually separated in 3D space due to their large difference in depth). Some depth artifacts noticeable are attributed to the insufficient scene overlapping between adjacent camera frames because $\mu$FTP relies on certain spatio-temporal redundancy for phase unwrapping.

\section*{Discussion}

\noindent
In this study, we have demonstrated $\mu$FTP, which achieves dense and precise 3D shape measurement of complex scenes at 10,000 fps, with a depth accuracy better than 80 $\mu$m and a temporal uncertainties below 75 $\mu$m over a large measurement volume of 400 mm $\times$ 275 mm $\times$ 400 mm. $\mu$FTP has several advantages over conventional high-speed 3D imaging methods. Not only can the high-resolution, unambiguous depth information be retrieved from two images, but also high-quality 2D textures (white patterns) provided simultaneously with the 3D geometry. This allows filling in the speed gap between high-speed 2D photography and fast 3D sensing, pushing the speed limits of unambiguous, motion-artifact-free 3D imaging to the range of tens of kilo-hertz (half of the native camera frame rate). The effectiveness of $\mu$FTP has been verified by several experiments on various types of transient events, including objects that are rapidly moving or undergoing sudden shape deformation. Experimental results suggest the great potential of $\mu$FTP in a vast number of applications, such as solid mechanics, material science, fluid dynamics, and biomedical research. Furthermore, $\mu$FTP is highly flexible: the fringe pitches, number of wavelengths, sliding window length can be adjusted according to the surface characteristics and motion speed of objects, for instance using more than 3 wavelengths to achieve higher reconstruction reliability for more complex objects when the motion speed is lower than the camera frame rate.

Being a recording and post-processing technique, the processing speed of $\mu$FTP has not yet been fully optimized. We have implemented the $\mu$FTP reconstruction in MATLAB (the RGC algorithm is written in C++ language and called from MATLAB using the Mex ``MATLAB Executable" dynamically linked subroutine). The reconstruction code can be accessed from our website \cite{77}. The time required for reconstructing one 3D frame is approximately 870 ms on a desktop computer (Intel Core i7-4790 CPU 3.6 GHz, 16 GB RAM). The processing speed can be significantly improved by using graphics processing units (GPUs), as the involved algorithms, such as 2D fast Fourier transform and pixel-wise PDM phase unwrapping are highly parallelizable. This could further enable $\mu$FTP to execute real-time 3D video reconstruction with low latency, if a high-speed camera with synchronous data acquisition capability is employed. Moreover, due to the very similar architecture and general versatility, $\mu$FTP is possible to be extended to other computational illumination based imaging techniques for high-speed imaging tasks, such as structured illumination microscopy \cite{18,79,80} and computational ghost imaging \cite{81,82}.

\bibliography{References2}

\section*{Acknowledgements}

This work was supported by the National Natural Science Fund of China (61505081, 111574152), Final Assembly ‘13th Five-Year Plan’ Advanced Research Project of China (30102070102), ‘Six Talent Peaks’ project of Jiangsu Province, China (2015-DZXX-009), ‘333 Engineering’ Research Project of Jiangsu Province, China (BRA2016407, BRA2015294), Fundamental Research Funds for the Central Universities (30917011204, 30916011322). C. Zuo thanks the support of the ‘Zijin Star’ program of Nanjing University of Science and Technology.

\section*{Author contributions statement}

C.Z. and T.T. proposed the idea. C.Z., T.T. and S.F. performed experiments. C.Z. and T.T. analyzed the data. L.H. contributed to the system calibration. S.F. contributed to 3D point cloud visualization; C.Z. and Q.C. and A.A. conceived and supervised the research. All authors contributed to writing the manuscript.

\section*{Additional information}

\textbf{Competing financial interests:} The authors declare no competing financial interests.

\end{document}